\documentclass[twoside,english]{article}
\usepackage[T1]{fontenc}
\usepackage[utf8]{luainputenc}
\usepackage{geometry}
\usepackage{graphicx}
\usepackage[numbers]{natbib}

\makeatletter
\usepackage{spconf}

\usepackage{tikz}
\usepackage{pgfplots}
\usetikzlibrary{patterns}

\ninept
\name{Seyed Hamidreza Mohammadi
\thanks{This material is based upon work supported by the National Science Foundation under Grant No.~0964468.}}
\address{Center for Spoken Language Understanding, Oregon Health \& Science University\\
Portland, OR, USA\\
{\small \tt mohammah@ohsu.edu}}

\makeatother

\usepackage{babel}
\begin{document}

\title{Reducing one-to-many problem in Voice Conversion by \\equalizing
the formant locations using dynamic frequency warping}

\author{Seyed Hamidreza Mohammadi}
\maketitle
\begin{abstract}
In this study, we investigate a solution to reduce the effect of one-to-many
problem in voice conversion. One-to-many problem in VC happens when
two very similar speech segments in source speaker have corresponding
speech segments in target speaker that are not similar to each other.
As a result, the mapper function usually over-smoothes the generated
features in order to be similar to both target speech segments. In
this study, we propose to equalize the formant location of source-target
frame pairs using dynamic frequency warping in order to reduce the
complexity. After the conversion, another dynamic frequency warping
is further applied to reverse the effect of formant location equalization
during the training. The subjective experiments showed that the proposed
approach improves the speech quality significantly.

\end{abstract}

\section{Introduction}

The task of Voice Conversion (VC) is to convert speech from a source
speaker to sound similar to that of a target speaker's. Various approaches
have been proposed; most commonly, a generative approach analyzes
speech frame-by-frame and then maps extracted source speaker features
towards target speaker features, with a subsequent synthesis procedure.
The mapping is achieved using a non-linear regression function, which
must be trained on aligned source and target features from existing
parallel or artificially parallelized~\citep{erro2010inca} speech.
In a parallel speech corpus, the utterances for both speakers have
the same linguistic content.

In this study, we assume that a parallel speech corpus is available.
For each utterance pair, a time alignment is performed to equalize
the difference in speaking rate of the speakers. Helander et. al.~\citep{helander2008impact}
extensively study the impact of frame alignment on VC performance.
They compare various approaches to aligning speech such as hand-labeling,
Dynamic Time Warping (DTW), and using Automatic Speech Recognition
(ASR) output. They conclude that with using variations of DTW, they
could achieve the performance of hand-labeling alignment. The alignment
can be updated after one or several iterations of training~\citep{erro2010inca}.
The output of this stage is the source-target frame pairs that are
supposed the same speech content as each other. The mapping function
is usually trained on these aligned speech segments.

One inherent problem in VC is one-to-many problem~\citep{mouchtaris2007conditional}.
For two or more similar aligned speech segments, there is no guarantee
that there exist corresponding target speech segments that are similar.
In other words, the mapping problem that VC is trying to solve is
not a \textit{function}. This makes the learning the mapping function
a challenging task. With such frame pairs present, some methods such
as Gaussian Mixture Models (GMMs)~\citep{kain1998spectral,stylianou1998continuous}
or Neural Networks~\citep{desai2009voice,mohammadi2014voice} might
result in over-smoothing or a muffling effect in the converted speech.
The reason is that the mapper is trying to learn a mapping that does
not have a function property. Since these mappers usually try to find
a solution with most similarity to the target speech typically by
trying to find a minimum mean squared error (MMSE) solution, the one-to-many
problem causes these mappers to converge to a solution that is most
similar to all the dissimilar target segments. This will ultimately
contribute to producing average speech segments that have over-smoothing
or muffling property. Some approaches try to reduce the muffling property
by applying a post-processing to match the variance of the converted
speech features to the original target features~\citep{toda2005spectral},
but it does not address this inherent problem in VC. The effect of
one-to-many problem might be different in other VC approaches. For
example, in approaches such a codebook~\citep{abe1988voice,arslan1999speaker}
or dictionary mapping~\citep{sundermann2006text,wu2013exemplar},
it might result in discontinuous speech, since the similar source
segments are used for finding target speech segments which are dissimilar.
Concatenating these dissimilar segments results in audible discontinuities
in converted speech.

Various factors might be the cause of the one-to-many problem. One
reason might be that people utter certain speech segments different
from each other because they pronounce words in the same context different
from each other. As a result, the source speaker might say a word
in two different sentences similarly, but the target speaker pronounces
the word differently. Another reason for the one-to-many problem might
be due to the rendition differences. When one person utters a certain
sentence multiple times, there is difference between the multiple
renditions of the sentence. This has been shown by objective measure
differences between the sentences uttered by the same speaker~\citep{kain2001high}.
This is another contribution to the one-to-many problem.  

\begin{figure*}
\begin{centering}
\includegraphics[scale=0.7]{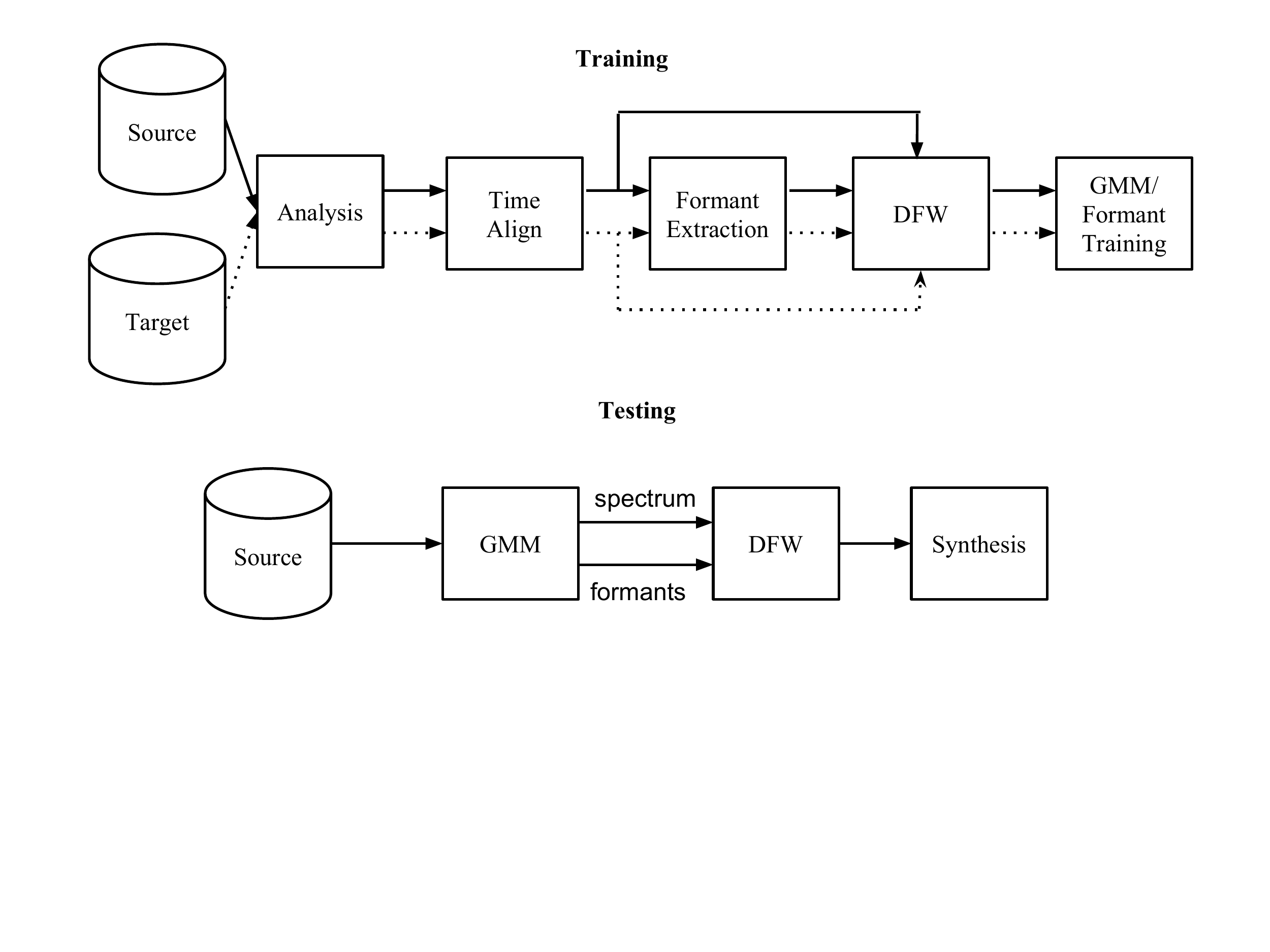}
\par\end{centering}

\vspace{-135bp}

\protect\caption{Proposed Framework\label{fig:Proposed-Framework}}
\end{figure*}

Mouchatris et. al.~\citep{mouchtaris2007conditional} have studied
one-to-many problem by extending the vector quantization (VQ) VC technique
to conditional VQ, which can capture one-to-many relationships. It
uses hard clustering on source frames independently from target frames.
Godoy et. al.~\citep{godoy2009alleviating} proposed to solve this
problem by considering context-dependent information. They consider
phonetic information in GMM framework. They also studied the effect
of training only on source versus training on the joint source-target
space and isolating one-to-many mappings from training using a threshold.
Turk et. al.~\citep{turk2006robust} also proposes to filter out
some source-target pairs that are unreliable based on some type of
confidence measure. 

\begin{figure}
\begin{centering}
\includegraphics[scale=0.33]{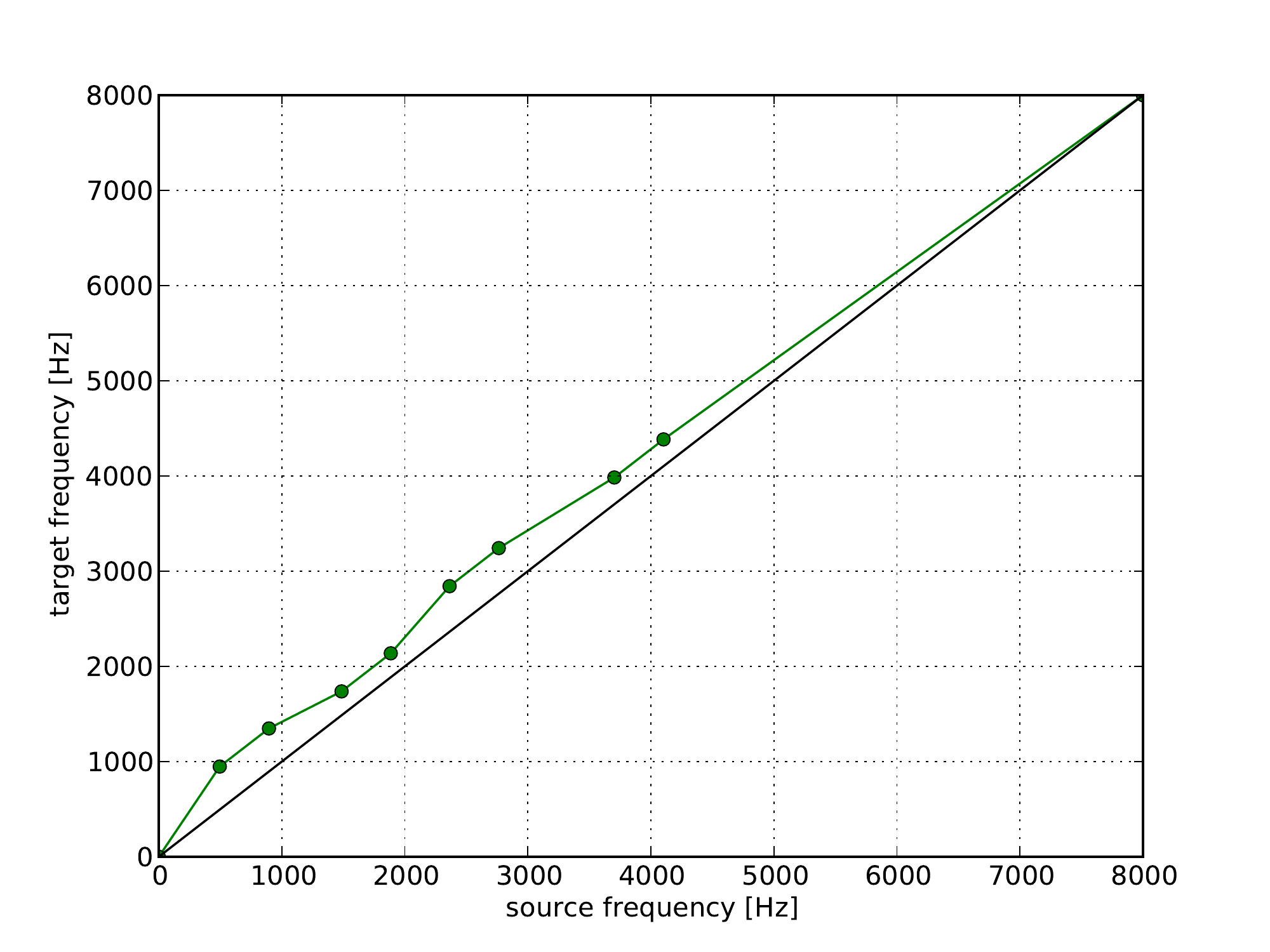}
\par\end{centering}

\protect\caption{Dynamic frequency warping\label{fig:Dynamic-frequency-warping}}
\end{figure}

In this study, we posit that one cause of the one-to-many problem
is the difference in formant locations. The formant difference in
target speech segments is a major factor that causes widening in synthesized
formants, since the mapping tries to be as similar to the two different
target speech segments. We propose to equalize the formant locations
of the source and target speech by warping the target speech spectra
to match the source formant locations. The mapping function is trained
on the formant equalized speech. At conversion time, the source is
converted to formant normalized target. The synthesized speech is
then warped using a DFW VC approach to match the formants of the target
speaker. The overall proposed framework is described in Figure~\ref{fig:Proposed-Framework}.

The formant equalization is described in Section~\ref{sec:Formant-Equalization}.
The mapping function is described in Section~\ref{sec:Mapping-Function}.
We then detail our voice conversion experiments, including system
configurations and their objective and subjective evaluation, in Section~\ref{sec:Experiment}.
Finally, we conclude in Section~\ref{sec:conc}.

\section{Formant Equalization}

\label{sec:Formant-Equalization}
\begin{figure*}
\hspace*{0.5cm}\includegraphics[scale=0.3]{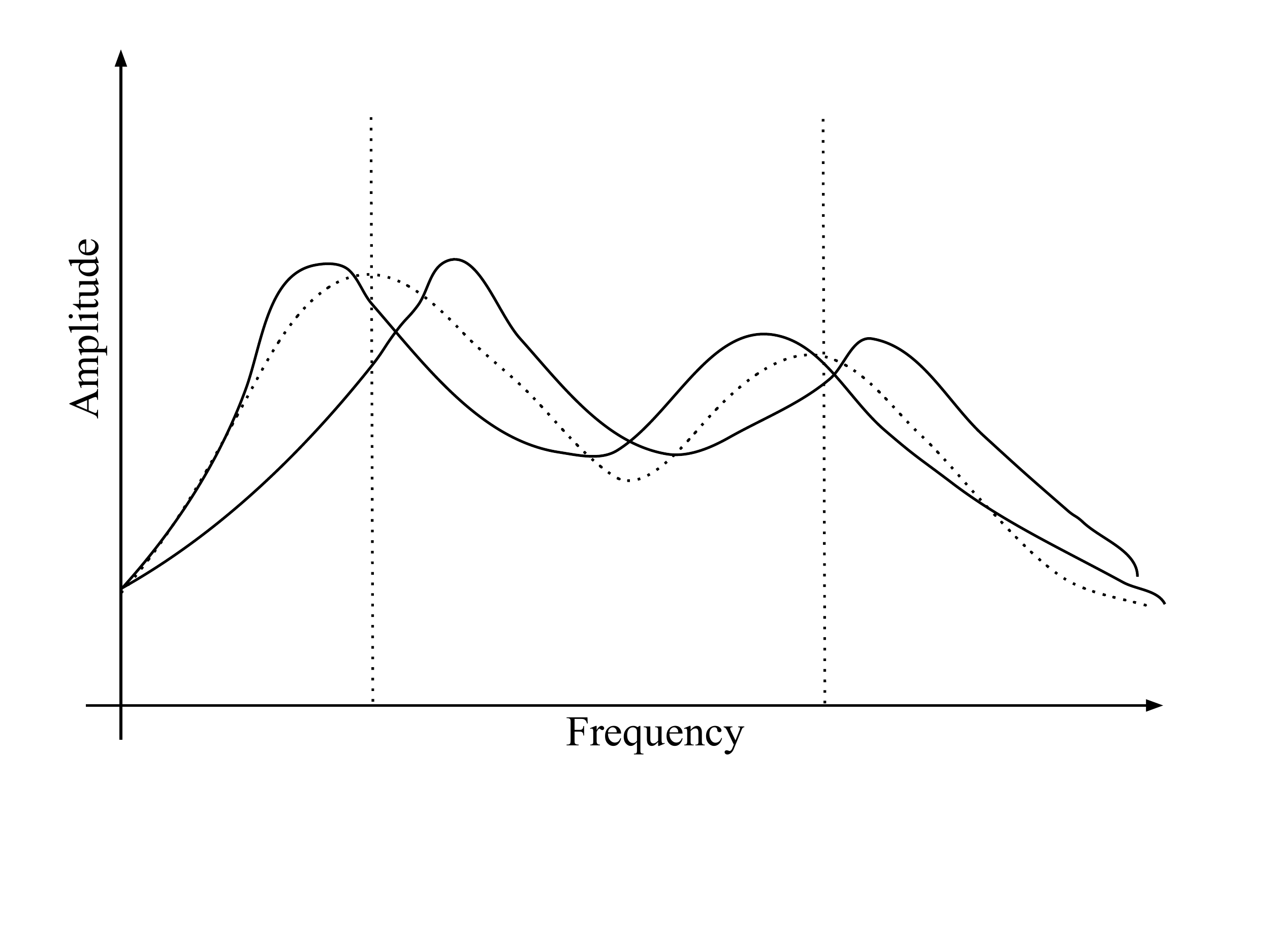}\hspace*{2cm}\includegraphics[scale=0.3]{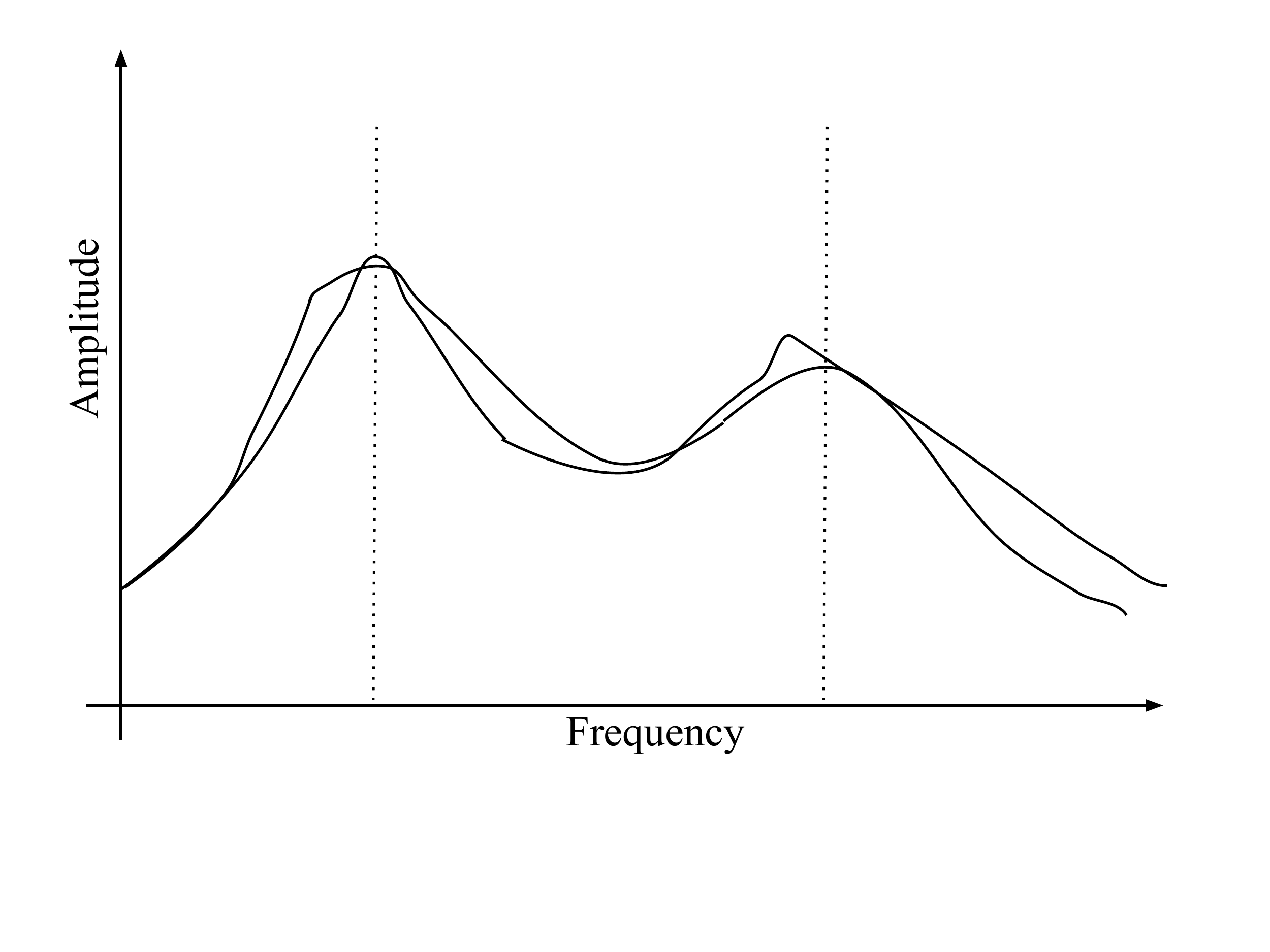}

\vspace{-35bp}

\protect\caption{Formant mismatch in target spectra for two similar source spectra
(left) and the proposed solution (right)\label{fig:one-to-many-problem-(left)}}
\end{figure*}

An example of a one-to-many problem
is depicted in Figure~\ref{fig:one-to-many-problem-(left)}. As can
be seen, two similar source spectra correspond to two dissimilar target
spectra. Various factors might contribute to this one-to-many problem.
One reason might be different contexts that different people say a
certain speech segments. For example, they might put different stress
on a specific word, or have different accents, or other high-level
reasons. Another reason that seems to might cause this one-to-many
problem might be due to differences in renditions. Even if one person
says the same sentence twice, there is a difference between the two
renditions of the sentence~\citep{kain2001high}. We posit that one
difference between the two spectra might be due to different formant
locations. This seems to be the most common symptom of the one-to-many
problem, and it is evident in Figure~\ref{fig:one-to-many-problem-(left)}.
We propose to normalize these effects by equalizing formant locations.

We use a method similar to DFW to equalize formants~\citep{erro2007weighted,erro2010voice,godoy2011spectral,mohammadi2013transmutative}.
First, the formant location and bandwidths are extracted from all
of the utterances using a signal processing algorithm. The utterances
are time-aligned using DTW. The aligned source-target formant information
is used as cues in DFW algorithm. Let the aligned feature sequence
be represented by $\mathbf{M}=[m_{1},m_{2},...m_{N}]$, the corresponding
log-spectrum by $S=[s_{1},s_{2},...s_{N}]$, the formant locations
by $F=[f_{1},f_{2},...f_{N}]$ and the formant bandwidth by $B=[b_{1},b_{2},...b_{N}]$.
The spectra are formant-equalized frame-by-frame. The formant of the
target spectrum are equalized to the source spectrum using

\begin{equation}
\bar{s}_{i}^{y}=dfw(s_{i}^{y},W(f_{i}^{y},b_{i}^{y},f_{i}^{x},b_{i}^{x}))
\end{equation}

The normalized spectrum $\bar{s}_{i}^{y}$ is converted back to the
feature domain $\bar{m}_{i}^{y}$. The warping function $W()$ is
constructed from both source and target estimated formant location
and bandwidth~\citep{mohammadi2013transmutative}. A sample warping
function is shown in Figure~\ref{fig:Dynamic-frequency-warping}
where the warping points are determined from source and target formant
location and bandwidth. If the warping causes the the target spectra
to be more different to source compared to no warping, we consider
this a sign of formant error and remove those frames.

\section{Mapping Function}

\label{sec:Mapping-Function}In this section, we briefly overview
the GMM mapping function~\citep{kain1998spectral}. Let $Z=[M^{x},\bar{M}^{y}]$
is the joint source-target spectral vector. A GMM represents the distribution
using $Q$ multivariate Gaussian

\begin{equation}
P(z)=\sum_{q=1}^{Q}\alpha N(z;\mu_{q},\Sigma_{q})
\end{equation}
where $N()$ is a normal distribution with $\alpha_{q},$ $\mu_{q}$
and $\Sigma_{q}$ as prior probability, mean and covariance of component
$q$, respectively. Each component would ideally represent an acoustic
class. The parameters of the GMM are calculated using the Expectation
Maximization (EM) algorithm on the joint vector $Z$. During conversion,
for each component, we estimate the MMSE of the target vector given
the source vector for each component

\begin{equation}
\hat{m}_{i}^{q}=E[M^{y}|M^{x}=m_{i}^{x}]=\mu_{q}^{y}-\Sigma_{q}^{xy}\Sigma_{q}^{xx-1}(m_{i}^{x}-\mu_{q}^{x})
\end{equation}
where each conversion in each component is weighted using the probability
that the frame $x_{i}$ belongs to the acoustic class described by
the component $q$

\begin{equation}
\hat{m}_{i}^{y}=\sum_{q=1}^{Q}\frac{\alpha_{q}N(m_{i}^{x};\mu_{q}^{x},\Sigma_{q}^{xx})}{\sum_{k=1}^{Q}\alpha_{k}N(m_{i}^{x};\mu_{k}^{x},\Sigma_{k}^{xx})}\hat{m}_{i}^{q}
\end{equation}

\section{Experiment}

\label{sec:Experiment}

\subsection{Configurations}

We use two male speakers from the CMU-arctic speech corpus~\citep{kominek2004cmu}.
We select RMS as source and BDL as target speaker. \textbackslash{}We
use 50 training sentences and 20 testing sentences from each speaker.
For analysis/synthesis, we use Ahocoder~\citep{erro2011improved},
which has shown good quality for parametric speech synthesis. We represent
spectrum using 39$th$-order MCEPs with $\alpha=0.42$ and 5$msec$
frame shifts, which are the recommended configurations for 16kHz waveforms.
We convert MCEP to log-spectrum and back for performing frequency-domain
warpings~\citep{erro2011improved}. We extract 4 formant location
and bandwidth using Snack~2.2 with LPC order 14~\citep{sjolander2000wavesurfer}.
We train a GMM with $Q=32$ components for transforming MCEPs. We
also train a separate GMM with $Q=8$ to map source and target formants
from source MCEPs.

\subsection{Objective Evaluation}

\label{sub:Objective-Evaluation}

\begin{figure*}
\begin{centering}
\includegraphics[scale=0.8]{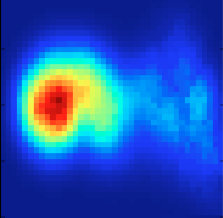} \hspace*{1cm}\includegraphics[scale=0.8]{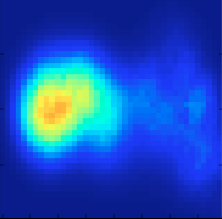}
\par\end{centering}

\protect\caption{2D visualization of the mapping complexity for the original feature
pairs (left) and the formant-equalized feature pairs (right)\label{fig:2d-visualization-of}}
\end{figure*}

This pre-processing step aims to make the mapping space less complex.
We compare the complexity map of the equalized and non-equalized formants
in Figure~\ref{fig:2d-visualization-of}. The approach to compute
the complexity map is to use a consistency measure based on the hypervolume
of the relative vectors in a certain region as represented by the
determinant of the data’s covariance matrix~\citep{mohammadi2012making}.
When vectors are mostly parallel in one region, the measure will have
a lower value (indicating relative consistency) than when relative
vectors are pointing in different directions (indicating relative
inconsistency). The weighted covariance for the region around $x'$
is given by

\begin{equation}
WeightedCov{}_{x'}=\frac{1}{{\scriptstyle \sum w_{i,x'}}}{\displaystyle \sum_{i=1}^{N}w_{i,x'}(y_{i}-\overline{y})(y_{i}-\overline{y})}\label{eq:weightcov}
\end{equation}
where the weights $w$ can be represented by any function that decreases
with the distance between $x$ and $x'$ . We chose the Gaussian function

\begin{equation}
w_{i,x'}=\exp(\frac{-\left\Vert x_{i}-x'\right\Vert ^{2}}{2\sigma^{2}})
\end{equation}
with $\sigma=0.1$. The final consistency measure was computed by
taking the determinant of the weighted covariance in Equation~\ref{eq:weightcov}.
For visualization purposes, the logarithm of the consistency value
is computed. As it is evident, the mapping shows less complexity when
the formants are equalized. This means for each source feature, there
are less dissimilar target features present. The 2-dimensional maps
are visualized by taking principal component analysis (PCA) in Figure~\ref{fig:2d-visualization-of}.
The raw speech feature pairs have a mel-cepstral distortion (melCD)
of 9.23dB and the the formant equalized version have a melCD of 8.38dB.

\subsection{Subjective Evaluation}

\label{sub:Subjective-Evaluation}To subjectively evaluate voice conversion
performance, we performed two perceptual tests: the first test measured
speech quality and the second test measured conversion accuracy (also
referred to as speaker similarity between conversion and target).
The listening experiments were carried out using Amazon Mechanical
Turk, with participants who had approval ratings of at least 90\%
and were located in North America. Both perceptual tests used three
trivial-to-judge sentence pairs, added to the experiment to filter
out any unreliable listeners. The statistical tests in this section
were performed using the Mann-Whitney test~\citep{mann1947test}.

\subsubsection{Speech Quality Test }

To evaluate the speech quality of the converted utterances, we conducted
a Comparative Mean Opinion Score (CMOS) test. In this test, listeners
heard two utterances A and B with the \emph{same} content and the
\emph{same} speaker but in two \emph{different} conditions, and are
then asked to indicate wether they thought B was better or worse than
A, using a five-point scale comprised of +2 (much better), +1 (somewhat
better), 0 (same), $-$1 (somewhat worse), $-$2 (much worse). It
is worthy to note that the two conditions to be compared differed
in exactly one aspect (either different mapping methods or different
number of training utterances). The experiment was administered to
20 listeners with each listener judging 20 sentence pairs.

Listeners' preference scores are shown in Figure~\ref{fig:subjective-similarity}.
FREQ (formant equalized) represents the proposed approach and ORIG
represents the baseline. The listeners preferred the speech quality
of the proposed framework. The improvement was shown to be significant.
The generated speech had less muffling effect which might be the reason
listeners judged those as higher quality.

\subsubsection{Conversion Accuracy Test}

To evaluate the conversion accuracy of the converted utterances, we
conducted a same-different speaker similarity test~\citep{Kain:01}.
In this test, listeners heard two stimuli A and B with \emph{different}
content, and were then asked to indicate wether they thought that
A and B were spoken by the \emph{same}, or by two \emph{different}
speakers, using a five-point scale comprised of +2 (definitely same),
+1 (probably same), 0 (unsure), $-$1 (probably different), and $-$2
(definitely different). One of the stimuli in each pair was created
by one of the four mapping methods, and the other stimulus was a purely
vocoded condition, used as the \emph{reference} speaker. The experiment
was administered to 20 listeners, with each listener judging 20 sentence
pairs.

\begin{figure}[t]
\begin{centering}
\begin{tikzpicture}
\begin{axis}[  
   width=4.5cm,
   height=5.2cm,
   ybar stacked, 
   enlargelimits=0.13,  
   symbolic x coords={tool1, tool2, tool3, tool4, 
	 tool5, tool6, tool7, tool8},  
   xtick=data, 
   ymax = 4,
   ymin=0,
   x tick label style={rotate=45,anchor=east}, 
   xticklabels={,,,,,,,},
   yticklabels={,,},
   bar width=25,
   ] 
\addplot+[black, ybar, fill=white] plot coordinates {(tool1,1.53) }; 
\addplot+[black, ybar, fill=gray] plot coordinates {(tool1,2.47) }; 

\pgfplotsset{     
after end axis/.code={  
      \node[black] at (axis cs:tool1,0.5){\small{ORIG}};
      \node[white] at (axis cs:tool1,3.5){\small{FREQ}};

    } 
}

\end{axis} 

\end{tikzpicture}\usetikzlibrary{patterns}
\makeatletter 
\tikzset{nomorepostaction/.code=\let\tikz@postactions\pgfutil@empty} 
\makeatother
\begin{tikzpicture}

\begin{axis}[
    width=5.0cm,
    height=5.2cm,
    xtick={1,...,2},
	ymin=0.0,
	ymax=0.5,
    enlarge x limits=0.2,
    xticklabels={
    },    
    grid=major,
    bar width=25,
    ybar     
    ]
\addplot[
    fill=white,
    draw=black,
    point meta=y,
    every node near coord/.style={inner ysep=5pt},
    error bars/.cd,
        y dir=both,
        y explicit 
]  
table [y ] {
x   y           error    label

1   0.26    0.01 4
2   0.54    0.01 4

};
\addplot[     
    fill=gray,
    draw=black,
    every path/.style=
    {
         postaction={ 
                  nomorepostaction
         }     
    },
    point meta=y,
    every node near coord/.style={inner ysep=5pt},
    error bars/.cd,
        y dir=both,
        y explicit 
]  
table [y ] {
x   y           error    label 

1   0.28    0.01 4

};
\legend{ORIG, FREQ} 
\draw [dashed] ({rel axis cs:1.9,0.58}-|{axis cs:1.9,0.58}) -- ({rel axis cs:1.9,0.92}-|{axis cs:1.9,0.92}) -- ({rel axis cs:2.8,0.92}-|{axis cs:2.8,0.92}) -- ({rel axis cs:2.8,0.86}-|{axis cs:2.8,0.86});
\draw [dashed] ({rel axis cs:1.8,0.58}-|{axis cs:1.8,0.58}) -- ({rel axis cs:1.8,0.95}-|{axis cs:1.8,0.95}) -- ({rel axis cs:3.8,0.95}-|{axis cs:3.8,0.95}) -- ({rel axis cs:3.8,0.86}-|{axis cs:3.8,0.86});
\draw [dashed] ({rel axis cs:2.1,0.23}-|{axis cs:2.1,0.23}) -- ({rel axis cs:2.1,0.60}-|{axis cs:2.1,0.60}) -- ({rel axis cs:4.2,0.60}-|{axis cs:4.2,0.60}) -- ({rel axis cs:4.2,0.50}-|{axis cs:4.2,0.50});
\draw [dashed] ({rel axis cs:2.2,0.23}-|{axis cs:2.2,0.23}) -- ({rel axis cs:2.2,0.55}-|{axis cs:2.2,0.55}) -- ({rel axis cs:3.2,0.55}-|{axis cs:3.2,0.55}) -- ({rel axis cs:3.2,0.40}-|{axis cs:3.2,0.40});

\end{axis} 
\end{tikzpicture}
\par\end{centering}

\protect\caption{\label{fig:subjective-similarity}Speech quality (left), Conversion
accuracy (right)}
\end{figure}
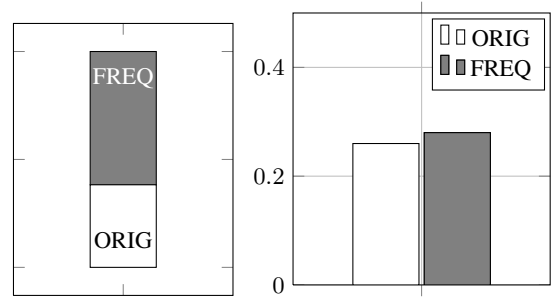

Listeners' average response scores are shown in Figure~\ref{fig:subjective-similarity}.
The difference between the two systems were not significant. This
was shown using a significance test. One reason might be using the
DFW directly on the log-spectrum domain, and also the formant estimation
mismatches that are inevitable. For controlling for the first problem,
using other warping approaches such as pole-shifting might be helpful.
For the second problem, hand-corrected formant values can be used
to see the effect of the proposed approach with the ground truth information
available to us.

\section{Conclusion}

\label{sec:conc}In this study, we investigated a solution to reduce
the effect of one-to-many problem in voice conversion. We proposed
to equalize the formant location of source-target frame pairs using
dynamic frequency warping in order to reduce the complexity. Finally,
A dynamic frequency warping is further applied after the conversion
to reverse the effect of formant location equalization. We were able
to show a significant gain in speech quality. Two issues present themselves
here. The issue is using DFW directly on the log-spectrum domain,
which might cause distorted-looking spectra, specially if there is
a formant error. For controlling for this problem, using other warping
approaches such as pole-shifting might be helpful. The other more
important problem is the formant estimation mismatches that are inevitable.
For solving this problem, hand-corrected formant values can be used
for experimentation purposes to see the real effect of the proposed
approach with the ground truth formant information.

\pagebreak{}

\bibliographystyle{IEEEbib}
\bibliography{references}

\end{document}